\documentclass[aps,prl,floatfix,twocolumn]{revtex4}
\usepackage{setspace}
\usepackage{graphics}
\usepackage{dcolumn}
\usepackage{epsfig}

\usepackage{setspace}

\newcommand{\simlt}
      {\ifmmode       { \raisebox{-.8em}{$<$}\atop\sim}
         \else        {$\raisebox{-.8em}{$<$}\atop\sim$}
      \fi}

\begin{document}

\bibliographystyle{prsty}

\title
{Spin Polarization of Photoelectrons from Topological Insulators}
\author{Cheol-Hwan Park}
\author{Steven G. Louie}
\email{sglouie@berkeley.edu}
\affiliation{Department of Physics, University of California at Berkeley,
Berkeley, California 94720\\
Materials Sciences Division, Lawrence Berkeley National
Laboratory, Berkeley, California 94720}

\date{\today}

\begin{abstract}
We show that the degree of spin polarization
of photoelectrons from the surface states of topological insulators is 100~\%
if fully-polarized light is used as in typical photoemission measurements,
and hence can be significantly {\it higher} than that of the initial state.
Further, the spin orientation of these photoelectrons
in general can also be very different from that of the initial surface state
and is controlled by the photon polarization;
a rich set of predicted phenomena have recently been confirmed
by spin- and angle-resolved photoemission experiments.
\end{abstract}
\maketitle

Three-dimensional topological insulators (TIs) are
strong spin-orbit interaction materials characterized by
a bulk electronic gap and metallic topological surface state (TSS)
bands with linear energy dispersions~\cite{PhysRevLett.98.106803,PhysRevB.75.121306,PhysRevB.79.195322}.
The predicted linear energy dispersion of the TSSs in TIs were first
observed in angle-resolved photoemission spectroscopy (ARPES)
measurements~\cite{hsieh_nature2008}.
TIs are considered to be a promising candidate for
spintronic devices because of their spin-momentum
locking~\cite{zhang:natphys,PhysRevLett.105.266806,RevModPhys.82.3045}.
Aspects of the spin distribution of the TSSs in TIs
have been measured by spin-resolved ARPES
experiments~\cite{hsieh_science2009,hsieh_nature2009,PhysRevLett.106.216803,arXiv:1101.3985,PhysRevLett.106.257004,PhysRevB.84.165113}.

Because the spin orientation of the photoelectron in the specific conditions
used in previous spin-resolved ARPES studies agreed with the expected
picture of the spin distribution of the TSS electrons in TIs [Fig.~1(a)],
photoemission matrix element effects were
neglected in analyzing the spin polarization of the photoelectrons.
(On the other hand, matrix element effects have been used in analyzing the
circular dichroism of the TSS electrons in a
TI~\cite{PhysRevLett.107.077601,PhysRevLett.107.207602,PhysRevLett.108.046805,arxiv:1108.1053}.)

Here, we find that
the degree of spin polarization of photo-ejected electrons
as defined in typical measurements with fully-polarized light
is significantly higher
(can in principle be 100\%) than that of the TSSs ($\sim50$\%
for Bi$_2$Se$_3$ and Bi$_2$Te$_3$ according to
first-principles calculations~\cite{PhysRevLett.105.266806}),
explaining the origin of high values of the measured degree of
polarization ($75\%$ and $>85\%$ in Pan {\it et al.}~\cite{PhysRevLett.106.257004}
and Jozwiak {\it et al.}~\cite{PhysRevB.84.165113}, respectively,
from experiments on Bi$_2$Se$_3$).
Moreover, using the symmetries of the TI surface,
we find that electron-photon interactions
can completely alter the spin orientation
of the photo-ejected electrons relative to that of the initial state and that the spin
orientation of these photoelectrons can be controlled
via light polarization tuning.
For linearly polarized light,
the detected spin orientation is significantly altered except for the specific
case of the wavevector {\bf k} of an initial state
being parallel to the in-plane component of the
light polarization $\hat{\epsilon}$.
For {\bf k} orthogonal to $\hat{\epsilon}$,
the two spins are predicted to be antiparallel to each other.
Moreover, for in-plane circularly polarized light,
the spins of the photoelectrons
are oriented either completely parallel or completely antiparallel
to the surface normal
depending on the handedness of the circular polarization.

The measured degree of spin polarization
is defined through the relation
\begin{equation}
P_{\rm max}=\max_{\{\hat{t}\}}\frac{I_{\hat t}-I_{-{\hat t}}}{I_{\hat t}+I_{-{\hat t}}}\,,
\label{eq:pol}
\end{equation}
where $I_{\hat t}$ and $I_{-{\hat t}}$ are the intensities for
the electron spin being aligned and anti-aligned with ${\hat t}$,
respectively.  The unit vector ${\hat t}={\hat t}_{\rm max}$ which maximizes
Eq.~(\ref{eq:pol}) defines the direction of the electron spin polarization.
Hsieh {\it et al.}~\cite{hsieh_nature2009} reported in-plane
$P_{\rm max}$ [i.\,e.\,, restricting $\hat{t}$ in Eq.~(\ref{eq:pol})
in the surface plane]
to be $20\%$ for photoelectrons from the TSSs of Bi$_2$Se$_3$,
Souma {\it et al.}~\cite{PhysRevLett.106.216803}
and Xu {\it et al.}~\cite{arXiv:1101.3985} reported
$60\%$ for those of Bi$_2$Te$_3$, and
Pan {\it et al.}~\cite{PhysRevLett.106.257004} and
Jozwiak {\it et al.}~\cite{PhysRevB.84.165113} reported
$75\%$ and $>85\%$, respectively, for those of Bi$_2$Se$_3$.
On the other hand, the degree of in-plane spin polarization
of the TSSs for both Bi$_2$Se$_3$ and Bi$_2$Te$_3$
obtained from first-principles calculations is
$\sim50\%$~\cite{PhysRevLett.105.266806}.
It is puzzling that the measured spin polarization
of the photoelectrons~\cite{PhysRevLett.106.257004,PhysRevB.84.165113}
can be significantly {\it higher} than that of the
corresponding TSS
obtained from theory~\cite{PhysRevLett.105.266806}.
Due to, e.\,g.\,,
spin-independent
background signals, the measured degree of spin polarization
is expected to always be {lower} than that from calculation.

We find that the degree of spin polarization of photoelectrons from the TSSs
of a TI is 100\%.
To show this, we start from the general expression for photocurrent from a detector
(that selects a specific three-dimensional wavevector
and hence a specific energy) with the spin quantization axis aligned
with $\hat{t}$:
\begin{equation}
I_{\hat t}\propto\left|\sum_{\{f| E_f=E_i+h\nu\}}\left<\hat{t},{\bf R}_D\right|
\left.f\right>\left<f\right|H^{\rm int}\left|i\right>\right|^2\,,
\label{eq:detected}
\end{equation}
where
$\left|i\right>$ is the initial TSS,
$\left|f\right>$ the photo-excited state,
$E_i$ and $E_f$ their respective energies,
$h\nu$ the photon energy,
and
$H^{\rm int}$ the light-matter interaction Hamiltonian connecting the two states.
State $\left|\hat{t},{\bf R}_D\right>$ is the detected state:
(i) its spin part is the eigenstate of ${\bf s}\cdot\hat{t}$
with eigenvalue $+1$ (${\bf s}$
is the Pauli matrix vector for spin half) and (ii) its spatial part is localized
at ${\bf R}_D$, where the detector is, far away from the sample surface.
(Note that ${\bf R}_D$ is on the trajectory
of the final-state wavepacket.)
We dropped the obvious prefactor $\delta(E-E_i-h\nu)$ in front of the
right hand side of Eq.~(\ref{eq:detected}) due to the energy-resolving
detector which collects electrons with energy $E$.
A formal derivation of Eq.~(\ref{eq:detected}) is reserved for
interested readers~\cite{suppl}. Here we discuss the
physical meaning of Eq.~(\ref{eq:detected}).
First, we note that the detector reads spin character of the photo-excited state
at ${\bf R}_D$; hence,
the near-surface part of the wavefunction affects the measured spin only
indirectly through the matrix element $\left<f\right|H^{\rm int}\left|i\right>$.
Second, a summation of the transition amplitude over degenerate
photo-excited states $\left|f\right>$'s is necessary because,
even in principle, we cannot tell which $\left|f\right>$
is involved in the detection~\cite{feynman3}.

If we denote
\begin{equation}
\left|f'\right>=\sum_{\{f| E_f=E_i+h\nu\}}\left|f\right>\left<f\right|H^{\rm int}\left|i\right>\,,
\label{eq:ip}
\end{equation}
then we may rewrite Eq.~(\ref{eq:detected}) as
\begin{equation}
I_{\hat t}\propto\left|\left<\hat{t},{\bf R}_D\right|\left.f'\right>\right|^2\,.
\label{eq:detected2}
\end{equation}
Since the measurement is performed
at ${\bf R}_D$,
far away from the sample where $\left|f\right>$'s
are eigenstates of the free-electron Hamiltonian, the Bloch periodic part
of the state $\left|f'\right>$ can be regarded as a position-independent
two-element spinor in evaluating $I_{\hat{t}}$ by Eq.~(\ref{eq:detected2}).
Then,
from basic spin physics, we always can find a vector $\hat{t'}$
satisfying $\left|f'\right>$ being the eigenstate of ${\bf s}\cdot\hat{t'}$
with eigenvalue $+1$ and the eigenstate of ${\bf s}\cdot(-\hat{t'})$
with eigenvalue $-1$.
Obviously $I_{{\hat t'}}\neq0$ and $I_{-{\hat t'}}=0$ for this
particular orientation and we have
$P_{\rm max}=100\%$ from Eq.~(\ref{eq:pol}),
i.\,e.\,, the degree of spin polarization of photo-ejected electrons
is always 100\% regardless of that of the initial electronic state.

An important ingredient that led us to this result is that the initial TSS
electronic state $\left|i\right>$ is not degenerate.
If it is, a hole will be left
in one of those {\it different} degenerate
TSSs $\left|i\right>$'s
after photodetection. Therefore, we can distinguish in principle which
initial TSS is involved in the measurement;
the detection probability amplitude corresponding to each $\left|i\right>$ should be first squared
and then summed, and not the other way round~\cite{feynman3}, i.\,e.\,,
the photocurrent $I_{{\hat t}}$ will be the sum of contributions coming
from all the degenerate initial TSSs $\left|i\right>$'s.
If this happens,
$P_{\rm max}\neq100\,$\%.
The simplest example is,
for a normal spin-degenerate material,
$I_{{\hat t}}$ and $I_{-{\hat t}}$ will always be the same regardless of the choice
$\hat{t}$, making $P_{\rm max}=0$.

Now we apply this general consideration to the case of the TSSs of a TI.
According to recent {\it ab initio} calculations~\cite{PhysRevLett.105.266806},
the averaged degree of spin polarization of the TSSs,
\begin{equation}
P^{\rm TSS}_{\rm ave}=\left|
\left<\psi(n,{\bf k})\right|{\bf s}\left|\psi(n,{\bf k})\right>
\right|\,,
\label{eq:Pave}
\end{equation}
where $\left|\psi(n,{\bf k})\right>$ is the two-component spinor wavefunction,
is roughly 50\,\%.
Because $\left|\psi(n,{\bf k})\right>$ is not an eigenstate of
a spin operator $\hat{t}\cdot{\bf s}$
for any
$\hat{t}$, the degree of spin polarization
for the TSSs had
to be defined as an averaged quantity.
On the contrary, the degree of spin polarization of the photoelectrons
$P_{\rm max}$ [Eq.~(\ref{eq:pol})] is 100\,\%
and that is so even if we take into account the imaginary part of the self energies
of the involved electronic states because
the detector probes the spin character (of the photoexcited states)
at ${\bf R}_D$ instead of taking its average over the entire space.
If the imaginary part of the initial-state self energy is considered,
the condition that ``the initial electronic state is not degenerate'' is naturally
modified to ``there are no other initial electronic states in the energy window
within the width of the imaginary part of the initial-state self energy.''

This result tells us that a direct comparison between
$P_{\rm max}$ [Eq.~(\ref{eq:pol})] and $P^{\rm TSS}_{\rm ave}$ [Eq.~(\ref{eq:Pave})]
is not meaningful, and solves
the apparent puzzle that the former from
experiments~\cite{PhysRevLett.106.257004,PhysRevB.84.165113}
is higher than the latter from theory ($\sim50$\,\%)~\cite{PhysRevLett.105.266806}.
Although the predicted degree of spin polarization of the photoelectrons from the TSSs is
100\%, the measured value will always be lower than this
due, e.\,g.\,, to
spin-unpolarized background signals and
finite resolution of the apparatus.

Equation~(\ref{eq:detected}) and our discussion based on it
are not confined to topological insulators and
can provide a guidance for the interpretation
of any spin-resolved photoemission experiment using fully-polarized light.

\begin{figure*}
\begin{center}
\includegraphics[width=1.8\columnwidth]{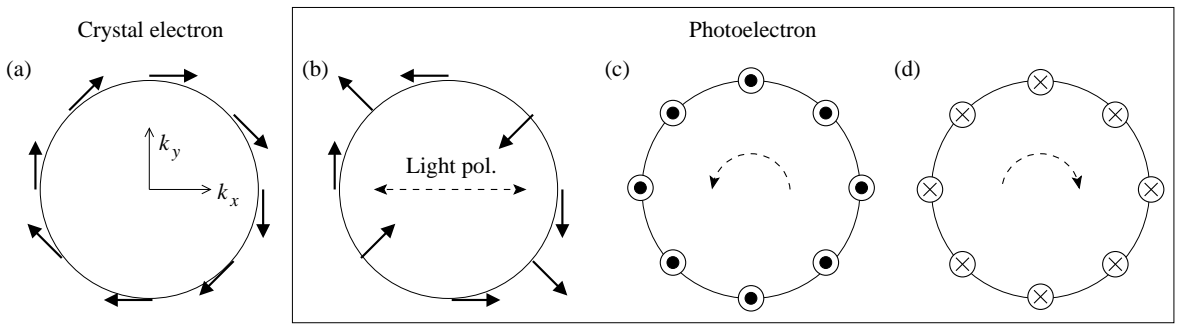}
\end{center}
\caption{(a) The spin orientation of the surface states in a topological insulator on an
    equi-energy contour in the upper band [$n=+1$ in Eqs.~(\ref{eq:E_TI})
    and~(\ref{eq:wfn_TI})].
    (b)-(d): Similar quantities as in (a) for the measured photoelectrons with light
    having linear (along the $+x$ direction), left-handed circular, and right-handed circular
    polarizations, respectively. (For the latter two, the handedness
    is defined from the viewpoint of the light source.)
    The surface normal direction is $\hat{z}$.}
\label{Fig1}
\end{figure*}

So far, we have discussed the magnitude of the spin polarization, without resorting
to the details of the system. Now,
by obtaining the state $\left|f'\right>$ in Eq.~(\ref{eq:ip})
using the specific symmetries of the TI surface, we determine the orientation of the spin polarization.
We adopt the commonly used effective Hamiltonian $H^0_{\rm TI}({\bf k})$ for
the Bloch periodic part $\left|\phi(n,{\bf k})\right>$ of the wavefunction of TSSs in a TI
with (in-plane) Bloch wavevector
${\bf k}=k_x\hat{x}+k_y\hat{y}$
($\hat{z}$ is along the surface normal) given by
\begin{equation}
H^0_{\rm TI}({\bf k})=\hbar v\,k\,(\sin\,\theta_{\bf k}\,\sigma_x-
\cos\,\theta_{\bf k}\,\sigma_y)\,,
\label{eq:H_TI_0}
\end{equation}
where $v$ is the band velocity,
$\theta_{\bf k}$ the angle between {\bf k} and the $+k_x$ direction, and
$\sigma_x$ and $\sigma_y$ are the Pauli matrices
acting on a two-component wavefunction, the so-called pseudospins.
In constructing the effective Hamiltonian
in Eq.~(\ref{eq:H_TI_0}),
the basis states defining the
$\left(\begin{array}{c}1\\0\end{array}\right)$ and
$\left(\begin{array}{c}0\\1\end{array}\right)$ column vectors,
corresponding to pseudospin up and down states, are constructed
from the two degenerate states at ${\bf k}=0$.

Here, we consider a class of materials having
the symmetry of the surface of Bi$_2$Se$_3$ or Bi$_2$Te$_3$;
however, the development can straightforwardly be
extended to other classes.
The two states at ${\bf k}=0$ which are used as basis,
$\left|\phi_{1}\right>$ and 
$\left|\phi_{2}\right>$,
can then be uniquely fixed by using the symmetry
properties:
\begin{eqnarray}
\Theta\left|\phi_{1}\right>=-\left|\phi_{2}\right>\,&,&\,\,\,
\Theta\left|\phi_{2}\right>=\left|\phi_{1}\right>\,,\nonumber\\
M\left|\phi_{1}\right>=i\left|\phi_{2}\right>\,&,&\,\,\,
M\left|\phi_{2}\right>=i\left|\phi_{1}\right>\,,\nonumber\\
C_3\left|\phi_{1}\right>=e^{-i\pi/3}\left|\phi_{1}\right>\,&,&\,\,\,
C_3\left|\phi_{2}\right>=e^{+i\pi/3}\left|\phi_{2}\right>\,,
\label{eq:sym}
\end{eqnarray}
where $\Theta$ is the time-reversal operator, $M$ the reflection operator,
$x\to-x$ ($\hat x$ is along the $\Gamma$K direction),
and $C_3$ the operator for ${2\pi}/{3}$ rotation around the $z$ axis.
The eigenvalue and Bloch periodic eigenstate are
\begin{equation}
E(n,{\bf k})=n\,\hbar\, v\,k\,,
\label{eq:E_TI}
\end{equation}
and
\begin{equation}
\left|\phi(n,{\bf k})\right>=
\frac{1}{\sqrt{2}}\left(
\left|\phi_1\right>
-n\,i\,e^{i\theta_{\bf k}}\left|\phi_2\right>
\right)=\left(
\begin{array}{c}
1\\
-n\,i\,e^{i\theta_{\bf k}}
\end{array}
\right)\,,
\label{eq:wfn_TI}
\end{equation}
respectively, where $n=\pm1$ is the band index.
The pseudospin expectation value of the TSS
[Eq.~(\ref{eq:wfn_TI})] is thus given by
\begin{equation}
\left<{\vec \sigma}\right>=n\left(\sin\theta_{\bf k}\,\hat{x}
-\cos\theta_{\bf k}\,\hat{y}\right)\,.
\label{eq:sigma0}
\end{equation}
It is known that the actual
spin expectation value $\left<{\bf s}\right>$
is aligned with the pseudospin expectation
value~\cite{RevModPhys.82.3045},
i.\,e.\,,
\begin{equation}
\left<{\bf s}\right>\propto n\left(\sin\theta_{\bf k}\,\hat{x}
-\cos\theta_{\bf k}\,\hat{y}\right)\,.
\label{eq:s0}
\end{equation}
The orientation of the spinor eigenstates in the upper band ($n=+1$)
is shown in Fig.~1(a).

In calculating photoemission matrix elements, we use the theoretical
framework of Wang {\it et al.}~\cite{PhysRevLett.107.207602}.
At small {\bf k},
it is assumed that
the final photoemission states $\left|f\right>$'s
are spin-degenerate because they are well within the
spin-degenerate bulk band continuum.
In matrix element calculations for small {\bf k},
we will approximate
the periodic part of the final state Bloch wavefunctions
$\left|\phi^f_\uparrow({\bf k},k_\perp)\right>$
and
$\left|\phi^f_\downarrow({\bf k},k_\perp)\right>$
by those with ${\bf k}=0$
and $k_\perp=\sqrt{ \frac{2m_e}{\hbar^2} (h\nu-E_{\rm D})}$,
where $k_\perp$ is the surface normal component
of the photoelectron wavevector,
$m_e$ the electron mass and $E_{\rm D}$ the energy of two-fold degenerate TSSs at ${\bf k}=0$.
We denote these two ({\bf k}-independent) states by
$\left|\phi^f_\uparrow\right>$
and
$\left|\phi^f_\downarrow\right>$,
respectively,
and use the same symmetry relations as in Eq.~(\ref{eq:sym})
to define
these states.
Also, we will neglect the momentum dependence of the velocity operator
${\bf v}({\bf k})=e^{-i{\bf r}\cdot{\bf k}}\,{\bf v}\,e^{i{\bf r}\cdot{\bf k}}$
[see Eq.~(\ref{eq:v})].
that has to be used in calculating the optical transition matrix element
between the periodic parts of the Bloch
wavefunctions.
This theoretical setup~\cite{PhysRevLett.107.207602} was employed to
find the energy- and momentum-dependent spin polarization of the
TSS in Bi$_2$Se$_3$ from time-of-flight ARPES measurement
with laser of energy 6.2~eV.
The spin polarization of TSS electrons thus obtained is in excellent
agreement with theory~\cite{PhysRevLett.103.266801}
and other spin-resolved ARPES
experiments~\cite{hsieh_science2009,PhysRevLett.106.216803,arXiv:1101.3985,PhysRevLett.106.257004}.
Therefore, our theoretical development
and predictions below should be valid when low-energy
light source is employed.

An important point to note here is that
$\left|\phi^f_\uparrow\right>$ and $\left|\phi^f_\downarrow\right>$
are {\it the actual spin-up and spin-down states along $\hat{z}$
far away from the surface in vacuum where the measurement is performed},
because we have imposed the symmetry constraints
of the system in Eq.~(\ref{eq:sym})~\cite{zhang:natphys,PhysRevLett.103.266801}.
Even though the real spin character of those two basis states
at the surface can be very complicated,
we can regard the chosen degenerate doublet
$\left|\phi^f_\uparrow\right>$ and $\left|\phi^f_\downarrow\right>$
the actual spin-up and spin-down states, respectively,
from a measurement point of view.

Next, the microscopic Hamiltonian $H_{\rm D}$
of an electron with spin-orbit coupling~\cite{PhysRev.100.580} is given by
\begin{equation}
H_{\rm D}=\frac{{\bf p}^2}{2m_e}+V+\frac{\hbar}{4m_e^2c^2}
\left(\nabla V\times{\bf p}\right)\cdot{\bf s}\,,
\label{eq:H_D}
\end{equation}
where $V$ is the one-electron potential.
Using Peierls substitution ${\bf p}\to{\bf p}-\frac{e}{c}{\bf A}$,
where ${\bf A}$ is the vector potential of the electromagnetic wave,
and the relation
$H^{\rm int}_{\rm D}({\bf A}) = H_{\rm D}\left({\bf p}-\frac{e}{\hbar
  c}{\bf A}\right)-H_{\rm D}({\bf p})$, we obtain
the electron-photon interaction Hamiltonian
\begin{equation}
H^{\rm int}_{\rm D}({\bf A})=-\frac{e}{c}{\bf A}\cdot{\bf v}\,,
\label{eq:Hint_D}
\end{equation}
where
\begin{equation}
{\bf v}=\frac{\bf p}{m_e}+
\frac{\hbar}{4m_e^2c^2}\left({\bf s}\times \nabla V\right)
\label{eq:v}
\end{equation}
is the velocity operator~\cite{PhysRev.100.580}.

We first consider the matrix elements
\begin{equation}
{\bf v}_{s,i}=\left<\phi^f_s\right|{\bf v}
\left|\phi_{i}\right>\,,
\label{eq:matrix_elt2}
\end{equation}
where $s=\uparrow$ or $\downarrow$
and $i=1$ or $2$
are the index of the photo-excite state and
the pseudospin basis index of the TSS state, respectively.
If we define
\begin{equation}
v_\pm=v_x\pm i\,v_y\,
\label{eq:vpm}
\end{equation}
and use the symmetry of the system
for both the initial and final states
[Eq.~(\ref{eq:sym})],
only the following four are non-zero
among twelve possible combinations~\cite{PhysRevLett.107.207602}:
\begin{eqnarray}
\left<\phi^f_\uparrow\right|v_+\left|\phi_2\right>
&=&
\left<\phi^f_\downarrow\right|v_-\left|\phi_1\right>^*
=
i\,\alpha\nonumber\\
\left<\phi^f_\uparrow\right|v_z\left|\phi_1\right>&=&
\left<\phi^f_\downarrow\right|v_z\left|\phi_2\right>
=i\,\beta\,,
\label{eq:matrix_elt3}
\end{eqnarray}
where $\alpha$ and $\beta$ are {\it real}
constants which can be determined from, e.\,g.\,,
first-principles calculations.

Plugging Eqs.~(\ref{eq:matrix_elt2}) and~(\ref{eq:matrix_elt3})
into Eq.~(\ref{eq:Hint_D}), we obtain the interaction Hamiltonian
matrix $H^{\rm int}_{\rm TI}({\bf A})$ connecting
the two basis functions of the TSSs,
$\left|\phi_1\right>$ and $\left|\phi_2\right>$,
to the spin-up and spin-down photoexcited states,
$\left|\phi^f_\uparrow\right>$ and $\left|\phi^f_\downarrow\right>$:
\begin{equation}
H^{\rm int}_{\rm TI}({\bf A}) = \frac{\alpha}{2c}\,e\,
\left[\left(A_y\,\sigma_x-A_x\,\sigma_y\right)
+i\,\left(\frac{2\beta}{\alpha}\right)\,A_z\,I
\right]\,,
\label{eq:H_TI_int_general}
\end{equation}
where $I$ is the $2\times2$ identity matrix.
In this study, we neglect the last term that depends on the {\it z}
component of the light polarization, i.\,e.\,, we set $A_z=0$,
in order to see the new physics clearly.
Because this term is proportional to the identity matrix,
it alone does not contribute to alteration of the
spin direction of a photoemitted electron from that of the initial state.

First we consider the case where the light is linearly polarized.
Then,
\begin{equation}
H^{\rm int}_{\rm TI}({\bf A}) = \frac{\alpha}{2c}\,eA\,
(\sin\,\theta_{\bf A}\,\sigma_x-\cos\,\theta_{\bf A}\,\sigma_y)\,,
\label{eq:H_TI_int}
\end{equation}
where $\theta_{\bf A}$ is the angle
between the vector potential {\bf A} and the $+x$ direction.
As we discussed before, the photocurrent is given by Eq.~(\ref{eq:detected}),
where in our case $\left|i\right>=\left|\phi(n,{\bf k})\right>$ [Eq.~(\ref{eq:wfn_TI})]
and $\left|f\right>$'s are $\left|\phi^f_\uparrow\right>$ and~$\left|\phi^f_\downarrow\right>$.
Since the photocurrent $I_{\hat{t}}$ is nothing but
the squared projection of the state
$\left|f'\right>$ in Eq.~(\ref{eq:ip}) to the detector state $\left|\hat{t},{\bf R}_D\right>$,
it is essential to know the spin orientation of $\left|f'\right>$.
Using Eq.~(\ref{eq:H_TI_int}),
we can write $\left|f'\right>$ [Eq.~(\ref{eq:ip})]
in the basis of $\left|\phi^f_\uparrow\right>$ and $\left|\phi^f_\downarrow\right>$ as
\begin{equation}
\left|\phi'(n,{\bf k})\right>=
\frac{1}{\sqrt{2}}
\left(
\begin{array}{c}
1\\
-n\,i\,e^{i(2\theta_{\bf A}-\theta_{\bf k})}
\end{array}
\right)\,.
\label{eq:wfn_TI2}
\end{equation}
Comparing
Eqs.~(\ref{eq:wfn_TI}) with~(\ref{eq:wfn_TI2}) and using Eqs.~(\ref{eq:sigma0})
and~(\ref{eq:s0}),
we find that the net effect of photoexcitation
on the detected electron spin polarization direction defined by
the direction $\hat{t}$ maximizing
$(I_{\hat{t}}-I_{-\hat{t}})/(I_{\hat{t}}+I_{-\hat{t}})$ in Eq.~(\ref{eq:pol})
is a rotation in direction through a change
\begin{equation}
\theta_{\bf k}\to
\theta_{\bf k}+2\Delta\theta_{{\bf A},{\bf k}}\,,
\label{eq:trans}
\end{equation}
where $\Delta\theta_{{\bf A},{\bf k}}\equiv\theta_{\bf A}-\theta_{\bf k}$
is the angle between the in-plane light polarization and the Bloch wavevector.
The spin expectation value of the photoelectron arriving at the detector
$\left<{\bf s}\right>_f$ in units of $\hbar/2$ is thus given by
\begin{equation}
\left<{\bf s}\right>_f=n\left[
\sin\left(\theta_{\bf k}+2\Delta\theta_{{\bf A},{\bf k}}\right)\hat{x}
-\cos\left(\theta_{\bf k}+2\Delta\theta_{{\bf A},{\bf k}}\right)\hat{y}
\right]\,.
\label{eq:s}
\end{equation}
(Note that the magnitude of the spin polarization
is 100\,\% in agreement with the above, general discussion.)

For the special case of $\theta_{\bf k}=\theta_{\bf A}$,
the spin orientation of the photoelectron [Eq.~(\ref{eq:s})]
is the same as that of the
initial TSS electron [Eq.~(\ref{eq:s0})].
However, in general, the two are different.
Especially, for
states whose Bloch wavevector {\bf k} is perpendicular to
the in-plane component of the light polarization {\bf A}
[i.\,e.\,, $\left|\Delta\theta_{{\bf A},{\bf k}}\right|=\pi/2$ in Eq.~(\ref{eq:trans})],
the initial and final spins are {\it antiparallel} to each other.

For in-plane circularly polarized light,
the vector potential is given by ${\bf A}=A(\hat{x}\pm\,i\,\hat{y})/\sqrt{2}$
with the $+$ and $-$ signs denoting
left-handed and right-handed circular polarizations
as defined from the viewpoint of the light source, respectively.
Then, within the effective Hamiltonian formalism [Eq.~(\ref{eq:H_TI_int_general})],
\begin{equation}
H^{\rm int}_{\rm TI}({\bf A}) = \frac{\alpha}{2\sqrt{2}c}\,eA\,(\pm\,i\,\sigma_x-\sigma_y)\,.
\label{eq:H_TI_int_circ}
\end{equation}
Applying the same argument as before, the spin polarization direction of
the photoelectron measured by a spin-resolved detector is that of
\begin{equation}
\left|\phi'(n,{\bf k})\right>_{\rm LHC}=
\left(
\begin{array}{c}
1\\
0
\end{array}
\right)
=\left|\phi^f_\uparrow\right>
\label{eq:wfn_TI2_cw}
\end{equation}
and
\begin{equation}
\left|\phi'(n,{\bf k})\right>_{\rm RHC}=
\left(
\begin{array}{c}
0\\
1
\end{array}
\right)
=\left|\phi^f_\downarrow\right>
\,,
\label{eq:wfn_TI2_ccw}
\end{equation}
for left-handed and right-handed circular polarizations, respectively.
The spin polarization direction of the photoelectrons are thus pointed along the
parallel ($+\hat{z}$) and antiparallel ($-\hat{z}$)
directions to the surface normal, for left-handed
[Fig.~1(c)] and right-handed [Fig.~1(d)] circular polarized light, respectively.

The phenomenon of photo-induced spin rotation in a TI
predicted here was not observed previously.
The reason is that $\Delta\theta_{{\bf A},{\bf k}}$ in Eq.~(\ref{eq:s})
was held fixed or was allowed to change little
in conventional spin-resolved ARPES measurements
by linearly polarized light
as different {\bf k} states were probed.
Recently, Lanzara and coworkers~\cite{jozwiak} have tuned $\Delta\theta_{{\bf A},{\bf k}}$
and confirmed the photo-induced spin rotations, as predicted here
for both linearly [Eq.~(\ref{eq:s})] and circularly
[Eqs.~(\ref{eq:wfn_TI2_cw}) and~(\ref{eq:wfn_TI2_ccw})]
polarized lights.

In conclusion,
we have shown that in spin-resolved
photoemission experiments, the measured degree of
spin polarization of photoelectrons with a specific energy
is 100\% under ideal condition
regardless of that of the initial state if the initial state is not degenerate
and fully-polarized light is used.
We have used this general principle to explain why the
degree of spin polarization of the photoelectrons from
the surface states of a topological insulator can be higher
than that of the initial states.
Using the specific symmetries of the system, we have further shown that
the spin polarization direction of photoexcited electrons
from the topological surface states
is in general very different from that of the initial states
and is dictated by the light polarization.
Our results provide
a theoretical basis for manipulation of the
spin polarization of the photoelectrons from
the topological surface states~\cite{jozwiak,PhysRevLett.105.057401,gedik:nnano}.

We thank A.\,Lanzara, C.\,Jozwiak, C.\,Hwang and J.\,D.\,Sau for
discussions. Theoretical part of this work was supported by National Science
Foundation Grant No. DMR10-1006184 and
simulations part by the Director,
Office of Science, Office of Basic Energy Sciences,
Materials Sciences and Engineering Division, U.S. Department of Energy under
Contract No. DE-AC02-05CH11231.

\section{Supplemental Material}

\subsection{1. Fermi's golden rule}

Consider an electronic system described by a single-particle Hamiltonian $H_0$
with energy eigenvalues and corresponding eigenstates $E_n$ and $\left|n\right>$,
respectively, where $n$ is the state index.
Suppose now that we have a time-dependent perturbation of the form
$H^{\rm int}(t)=H^{\rm int}({\bf A})\,e^{-2\pi i \nu t}$,
where ${\bf A}$ is the vector potential and $h\nu$ the photon energy,
in addition to the original Hamiltonian $H_0$. The total Hamiltonian is given by
\begin{equation}
H(t)=H_0+H^{\rm int}({\bf A})\,e^{-2\pi i\nu t}\,.
\label{eq:H}
\end{equation}
We are interested in the evolution of an electron which is at $\left|i\right>$
before turning on the time-dependent perturbation. Assuming that
the perturbation is turned on at $t=0$,
the wavefunction satisfying normalization condition
($\left<\psi(t)|\psi(t)\right>=1$) to first order can be written as
\begin{equation}
\left|\psi(t)\right>=e^{-iE_it/\hbar}\,\left|i\right>+\sum_f\,c_f(t)\,e^{-iE_ft/\hbar}\left|f\right>\,,
\label{eq:c_f}
\end{equation}
where $c_f(t)$ is the probability amplitude of finding an electron in state $\left|f\right>$
at a given time $t$. Obviously,
\begin{equation}
c_f(0)=0\,.
\label{eq:bc}
\end{equation}

Using Eqs.~(\ref{eq:H}), (\ref{eq:c_f}), and~(\ref{eq:bc}), we can solve the time-dependent
Schr\"odinger equation and arrive at
\begin{equation}
c_f(t)=-i\,e^{i\,t\,\frac{\Delta E_f}{2\hbar}}\,\frac{\sin{t\,\frac{\Delta E_f}{2\hbar}}}{\frac{\Delta E_f}{2}}\,
\left<f\right|\,H^{\rm int}\,\left|i\right>\,,
\label{eq:c_ft}
\end{equation}
where
\begin{equation}
\Delta E_f\equiv (E_f-E_i)-h\nu\,.
\label{eq:dE}
\end{equation}
Equation~(\ref{eq:c_ft}) leads to
\begin{equation}
P_f(t)\equiv|c_f(t)|^2=\frac{t}{\hbar^2}\,\frac{\sin^2{t\,\frac{\Delta E_f}{2\hbar}}}{t\,\left(\frac{\Delta E_f}{2\hbar}\right)^2}\,
\left|\left<f\right|\,H^{\rm int}\,\left|i\right>\right|^2\,.
\label{eq:c2_ft}
\end{equation}
Using the following relation
\begin{equation}
\lim_{a\to\infty}\frac{1}{\pi}\frac{\sin^2(ax)}{ax^2}=\delta(x)\,,
\label{eq:delta}
\end{equation}
we finally obtain
\begin{eqnarray}
\Gamma_{f\leftarrow i}&=&\lim_{t\to\infty}P_f(t)/t\nonumber\\
&=&\frac{\pi}{\hbar^2}\delta\left(\frac{\Delta E_f}{2\hbar}\right)
\left|\left<f\right|\,H^{\rm int}\,\left|i\right>\right|^2\nonumber\\
&=&\frac{2\pi}{\hbar}\delta(\Delta E_f)
\left|\left<f\right|\,H^{\rm int}\,\left|i\right>\right|^2\,,
\label{eq:goldenrule}
\end{eqnarray}
which is Fermi's golden rule.

\subsection{2. Spin-resolved photoemission experiment}

Now we find an expression for photocurrent $I_{\hat{t}}$ in spin-resolved photoemission experiment
with a detector aligned such that it captures electrons in a state
$\left|\hat{t},{\bf R}_D\right>$ of which
(i) the spin part is the eigenvector of ${\bf s}\cdot\hat{t}$
with eigenvalue $+1$ and (ii) the spatial part is localized at
the detector (${\bf r}={\bf R}_D$)
as in the main manuscript.
The photocurrent is then given by
\begin{equation}
I_{\hat{t}}\propto\lim_{t\to\infty}\left|\left<\hat{t},{\bf R}_D|\psi(t)\right>\right|^2/t\,.
\label{eq:Iph}
\end{equation}
Note that in the previous section, we have used
$\left|f\right>$ in the place of
$\left|\hat{t},{\bf R}_D\right>$ to derive Fermi's golden rule;
replacement of the final state is the only difference.

Using Eqs.~(\ref{eq:c_f}) and~(\ref{eq:c_ft}), we can write
$\left<\hat{t},{\bf R}_D|\psi(t)\right>$ as
\begin{widetext}
\begin{eqnarray}
\left<\hat{t},{\bf R}_D|\psi(t)\right>&=&
\sum_{E}\sum_{\{f|E_f=E\}}
-i\,e^{i\,t\,\frac{\Delta E_f}{2\hbar}}\,\frac{\sin{t\,\frac{\Delta E_f}{2\hbar}}}{\frac{\Delta E_f}{2}}\,
\left<f\right|\,H^{\rm int}\,\left|i\right>
\,e^{-iE_ft/\hbar}\,\left<\hat{t},{\bf R}_D|f\right>\nonumber\\
&=&
\sum_{E}
-i\,e^{i\,t\,\frac{\Delta E-2E}{2\hbar}}\,\frac{\sin{t\,\frac{\Delta E}{2\hbar}}}{\frac{\Delta E}{2}}\,
\sum_{\{f|E_f=E\}}
\,\left<\hat{t},{\bf R}_D|f\right>
\left<f\right|\,H^{\rm int}\,\left|i\right>\,,
\nonumber\\
\label{eq:step}
\end{eqnarray}
\end{widetext}
where $\Delta E=(E-E_i)-h\nu$.
An important knowledge we made use of in Eq.~(\ref{eq:step})
is that $\left<\hat{t},{\bf R}_D|i\right>=0$
because the initial state is localized at the crystal
and does not extend to the detector.
Obviously, a product of the terms in Eq.~(\ref{eq:step}) corresponding to different
$E$ values does not contribute to the photocurrent calculated from Eq.~(\ref{eq:Iph}).
Plugging Eq.~(\ref{eq:step}) into Eq.~(\ref{eq:Iph}) and using Eq.~(\ref{eq:delta})
again, we obtain the expression for the photocurrent:
\vspace{-0.7in}
\begin{widetext}
\begin{eqnarray}
I_{\hat{t}}&\propto&
\sum_{E}
\delta(E-E_i-h\nu)
\left|\sum_{\{f|E_f=E\}}\,
\left<\hat{t},{\bf R}_D|f\right>
\left<f\right|\,H^{\rm int}\,\left|i\right>\right|^2\nonumber\\
&\propto&
\left|\sum_{\{f|E_f=E_i+h\nu\}}\,
\left<\hat{t},{\bf R}_D|f\right>
\left<f\right|\,H^{\rm int}\,\left|i\right>\right|^2\,,
\label{eq:Iph2}
\end{eqnarray}
\end{widetext}
which is Eq.~(\ref{eq:detected}) of the main manuscript.
Equivalently, Eq.~(\ref{eq:Iph2}) can also be written as
\begin{widetext}
\begin{equation}
I_{\hat{t}}\propto
\left|\sum_f\, \delta(E_f-E_i-h\nu)\,
\left<\hat{t},{\bf R}_D|f\right>
\left<f\right|\,H^{\rm int}\,\left|i\right>\right|^2\,.
\label{eq:Iph3}
\end{equation}
\end{widetext}

\end{document}